\documentclass[conference]{IEEEtran}
\IEEEoverridecommandlockouts

\usepackage{amsmath}

\usepackage{graphicx}
\usepackage{csquotes}
\usepackage{algorithm2e}
\usepackage{paralist}
\usepackage[table,xcdraw]{xcolor}
\usepackage{balance}

\usepackage[section,numberedsection=autolabel]{glossaries}
\makeglossaries

\newglossaryentry{function}{name=function,%
    description={A single, self-contained piece of software, that performs a certain function}}

\newglossaryentry{feature}{name=feature,
    description={Composed of one or more functions connected together using a certain runtime environment, usually corresponds to a certain use-case}}

\newglossaryentry{runtime environment}{name=runtime environment,
    description={Communication middleware and virtualization mechanisms}}

\newglossaryentry{ECU}{name=ECU,
    description={Electronic Control Unit. It is an electronic device in a vehicle that is responsible for a single function}}

\newglossaryentry{TDD}{name=TDD,
    description={Test-Driven Development is a software development methodology that centers on the iterative creation of unit tests prior to the implementation of functional code~\cite{ref40:Beck2002} This approach mandates that a test case specifying the desired behavior of a code unit be written before the production code itself. As development progresses, the test suite continuously executes. New code is only written if it fulfills the requirements outlined in a failing test}}

\newglossaryentry{FDD}{name=FDD,
    description={Feature-Driven Development. It is a paradigm where the software system is iteratively developed in a series of steps, starting with an abstract model of the system, followed by extraction of a set of desired features, and ending with feature implementation and integration~\cite{ref41:Palmer2001}}}

\newglossaryentry{MBSE}{name=MBSE,
    description={Model-Based Systems Engineering is a formalized methodology within systems engineering that emphasizes using models as the primary means of information exchange and system representation~\cite{ref42:Incose2023}. This contrasts with traditional document-centric approaches. MBSE centers on creating and leveraging domain-specific models or metamodels, which capture system requirements, design, analysis, and verification elements throughout the development lifecycle}}

\newglossaryentry{contract}{name=contract,
    description={Design by contract is a software development methodology that emphasizes the explicit definition of formal contracts between software components~\cite{ref43:Mitchell2002}. These contracts specify preconditions (what must be true before a component is used), postconditions (what must be true after execution), and invariants (conditions that must always hold true). Design by contract can be enforced through runtime assertions, unit tests, or even integrated into a programming language's syntax. This approach enhances software reliability, eases debugging, and facilitates code comprehension}}

\newglossaryentry{metamodel}{name=Metamodel,
    description={Defines the language of system description by specifying abstract entities that are part of the system, a set of possible relations between them, and their attributes}}

\newglossaryentry{instance model}{name=Instance model,
    description={A model generated from the given metamodel, populated with actual objects with concrete attribute values; an implementation of the system described in the language of the metamodel}}

\newglossaryentry{OMG}{name=OMG,
    description={Object Management Group}}

\newglossaryentry{LLM}{name=LLM,
    description={Large Language Model}}

\newglossaryentry{OCL}{name=OCL,
    description={Object Constraint Language}}

\newglossaryentry{RACE}{name=RACE,
    description={Centralized Platform Computer Based Architecture for Automotive Applications}}

\newglossaryentry{Ecore}{name=Ecore,
    description={Language of the metamodel used in Eclipse Modeling Framework}}

\newglossaryentry{OEM}{name=OEM,
    description={Original Equipment Manufacturer}}

\makeatletter 
\newcommand{\linebreakand}{%
  \end{@IEEEauthorhalign}
  \hfill\mbox{}\par
  \mbox{}\hfill\begin{@IEEEauthorhalign}
}
\makeatother 

\title{LLM-based Iterative Approach to Metamodeling in Automotive \\\vspace*{20pt} \normalsize  \today{}}

\author{
\IEEEauthorblockN{Nenad Petrovic\IEEEauthorrefmark{1}}
\IEEEauthorblockA{email: nenad.petrovic@tum.de \\
orcid: 0000-0003-2264-7369}
\and
\IEEEauthorblockN{Fengjunjie Pan\IEEEauthorrefmark{1}}
\IEEEauthorblockA{email: f.pan@tum.de \\
orcid: 0009-0005-8303-1156}
\and
\IEEEauthorblockN{Vahid Zolfaghari\IEEEauthorrefmark{1}}
\IEEEauthorblockA{email: v.zolfaghari@tum.de \\
orcid: 0009-0004-0039-6014}
\and
\IEEEauthorblockN{Alois Knoll\IEEEauthorrefmark{1}}
\IEEEauthorblockA{email: k@tum.de \\
orcid: 0000-0003-4840-076X}
\linebreakand
\IEEEauthorrefmark{1}\IEEEauthorblockA{Technical University of Munich (TUM) \\
School of Computation, Information and Technology (CIT) \\
Chair of Robotics, Artificial Intelligence and Real-Time Systems}
}

\date{\today}

\begin{document}

\maketitle

\begin{abstract}
In this paper, we introduce an automated approach to domain-specific metamodel construction relying on Large Language Model (LLM). The main focus is adoption in automotive domain. As outcome, a prototype was implemented as web service using Python programming language, while OpenAI's GPT-4o was used as the underlying LLM. Based on the initial experiments, this approach successfully constructs Ecore metamodel based on set of automotive requirements and visualizes it making use of PlantUML notation, so human experts can provide feedback in order to refine the result. Finally, locally deployable solution is also considered, including the limitations and additional steps required.
\end{abstract}

\section{Introduction}
Evolution and paradigm shifts towards centralized vehicular systems in automotive have significant impact on related tools used for engineering and development as well \cite{refn1:Petrovic2024}. In automotive, model-driven engineering (MDE) tools and methods (such as SysML on top of AUTOSAR standard) are widely adopted in order to make the development more convenient for different experts involved into the process \cite{refn2:Schulze2012}. At core of model-driven tools, metamodels are used as formal representations capturing key domain concepts, their attributes and relations. Therefore, the change of automotive architectures implies update of the existing metamodels and their alignment with respect to current trends in order to support. However, this task is expensive and time-consuming, as it is usually done manually and requires sophisticated automotive domain expertise. Additionally, changes on metamodel-level might also require the updates when it comes to other software artifacts relying on its structure, such as model instance creation and visualization tools, especially when it comes to underlying parsers. Due to the previously mentioned facts, it can be observed that such changes lead to disruptions of the development process, making it less flexible and slower when it comes to innovation-oriented experimentation in this domain.

On the other side, the trending Large Language Models (LLMs) exhibit strong text summarization capabilities, making them potentially suitable for task that involve concept abstraction which is the case of metamodeling, which was approved in earlier research  \cite{refn3:Petrovic2023} \cite{refn31:Chen2024-1}. Therefore, in this paper, we examine the capability of state-of-art commercial LLM – OpenAI’s GPT-4.0o for metamodeling task, while the focus is on automotive domain. The goal of the proposed approach to automatically generate automotive metamodel based on OEM-provided requirements as input through multiple iterations, as LLM-based metamodeling approaches leveraging problem decomposition exhibit better results \cite{refn32:Chen2024-2}. For metamodel representation, we use widely adopted Eclipse Modeling Framework (EMF) and its Ecore format \cite{refn4:Ecore}, while PlantUML \cite{refn5:PlantUML} class diagram representation is used in order to visualize the intermediary result to the end-user. In order to demonstrate the methodology, a prototype web service is implemented using Python programming language. To the best of our knowledge, there aren't any similar tools targeting automotive domain available yet. 

\section{Implementation Overview}
\begin{figure*}[t]
    \centering
    \includegraphics[width=0.8\textwidth]{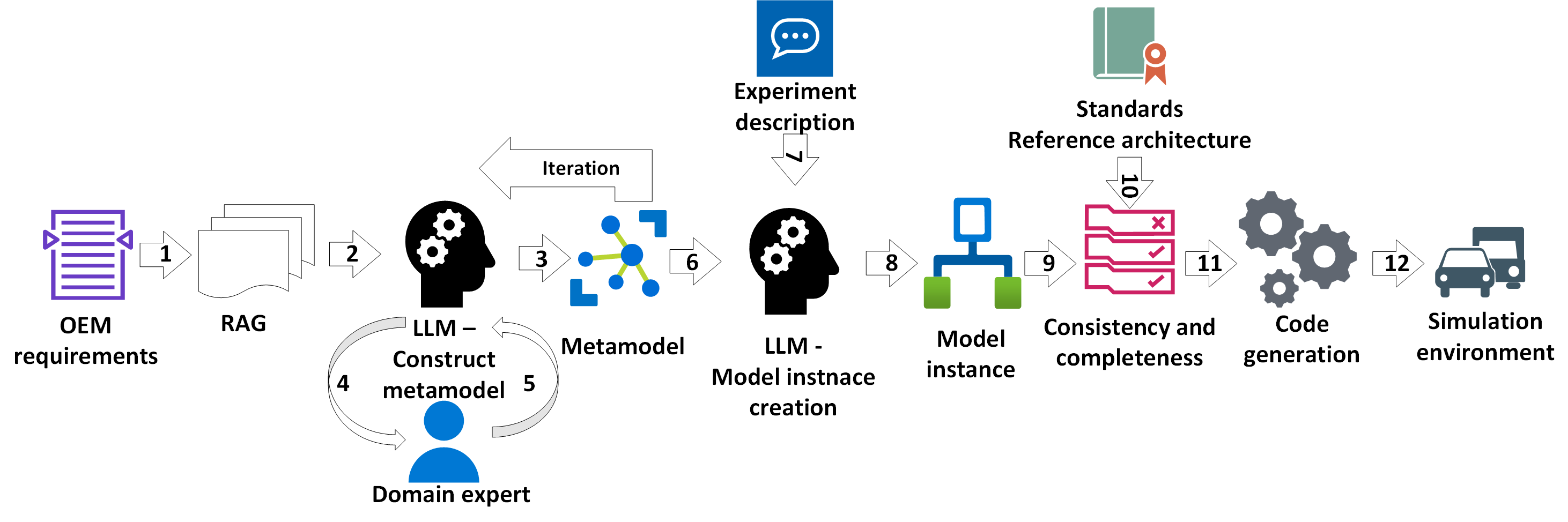}
    \caption{LLM-based workflow for iterative metamodeling in automotive domain: 1-Freeform text 2-Text chunks 3-Ecore metamodel (partial) and updated model parser 4-Metamodel visualization based on PlantUML 5-Expert feedback/update request 6-Ecore metamodel (complete) 7-Freeform text 8-Model instance 9-XMI file 10-OCL rules based on standards 11-Complete and consistent system model 12-Executable code}.
   \label{fign1}
\end{figure*}

In Fig. \ref{fign1}, a high-level depiction of the proposed method's workflow is given. Within the first step, automotive engineer provides a collection of requirements covering various aspects of the vehicular system. While the scope of the provided requirements can be arbitrarily large, it is expected to cover at least the aspects relevant to sensors and actuators of the vehicle, as they can be conveniently demonstrated in our simulation engine on top of CARLA, as presented in \cite{refn6:Petrovic2024-2}. 

In order to extract the requirements relevant to specific aspect of the vehicular system (such as filtering only sensor and actuator-related), we make use of Retrieval Augmented Generation (RAG) methodology, built upon our previous work from \cite{refn7:Zolfaghari2024}. RAG splits the provided requirements document into chunks, multiple smaller pieces which are leveraged in order to store and later retrieve information more efficiently from large input documents.

After that, LLM-based service is used to construct metamodel based on the provided set of requirements. As output of this process, a partial metamodel is constructed, which is not the final outcome, but intermediate product which is iteratively refined, based on additional requirements which are provided as input. In parallel, it is expected that LLM also generates updated model parser based on the current metamodel each time it is updated. The initial partial metamodel only contains "Vehicle" element which should be on top of hierarchy. Additionally, at each of the steps, the current state of the partial metamodel can be visualized as PlantUML class diagram, so human expert can have insight into its structure. Optionally, in case that further correction or refinement of the LLM-constructed metamodel is needed, human expert can provide feedback, which will be taken into account and could directly affect its structure. As already emphasized, the process of metamodel construction is iterative and its output is Ecore file, while PlantUML class diagram and PNG format image are constructed as side products intended for metamodel visualization. While PlantUML representation is created also using LLM, the image file is rendered using an auxiliary client library for Python - plantweb \cite{refn8:plantweb}.

In what follows, the underlying prompts used for metamodel construction are given. 

\textit{System prompt 1: "You are generating .ecore metamodel without additional comments."}

\textit{User prompt 1: "Based on description [requirements] update Ecore metamodel [current metamodel]".}

Additionally, PlantUML representation is also created for purpose of metamodel visualization based on the following prompts.

\textit{System prompt 2: "You are generating plantuml metamodel about vehicle with no additional comments. Subclass relations should be identified."}

\textit{User prompt 2: "Based on description [requirements] update plantuml class diagram [current plantuml]"}

Once metamodel construction process is finished through one or many iterations of adding new requirements, the produced Ecore file can be further used. There are two later steps where metamodel is required: 1) model instance creation - construction of XMI model instances using LLMs based on user-provided requirements with respect to given metamodel, as described in \cite{refn1:Petrovic2024}; 2) OCL rule generation - construction of design constraint rules based on reference architecture and standardisation-related documents, built upon works from \cite{refn9:Pan2024}. After those two steps, the model instance is checked in order to identify whether it is compliant to given set of OCL rules. The underlying mechanism is work in progress, while the initial results were presented in \cite{refn9:Pan2024}.

Finally, once the model instance is complete and contradiction-free, it can be further used for code generation. Based on our work from \cite{refn6:Petrovic2024-2}, we further generate code for CARLA simulator covering the aspects of vehicle configuration (sensors, actuators), environment (pedestrians, obstacles, weather conditions) and behavior (event chain for autonomous driving-related functionality, such as lane keeping and emergency braking).

The initial prototype is deployed on free version of Google Colab as Python web application relying on Flask with simple API, as shown in Table \ref{tn1}. In order to run it, about only 1.2GB of RAM was occupied, as it relies on cloud-based infrastructure provided by OpenAI for LLM execution. The first service endpoint is used for both creation of new and update of the existing metamodel, considering either vehicle requirements or human hints/feedback as input. The role of second service endpoint is to visualize the current state of the constructed metamodel.

\begin{table}[]
\caption{Prototype API overview.}
\begin{tabular}{|l|l|l|l|}
\hline
\begin{tabular}[c]{@{}l@{}}Service endpoint\end{tabular} & Method & Parameters & Output \\ \hline
\it{updateMetamodel} & POST & requirements & Ecore metamodel \\ \hline
\it{getCurrentMetamodel} & GET & - & PNG image \\ \hline
\end{tabular}
\label{tn1}
\end{table}

\section{Experiment scenario and evaluation}
\begin{figure*}[ht]
	\centering
	\includegraphics[width=0.8\textwidth]{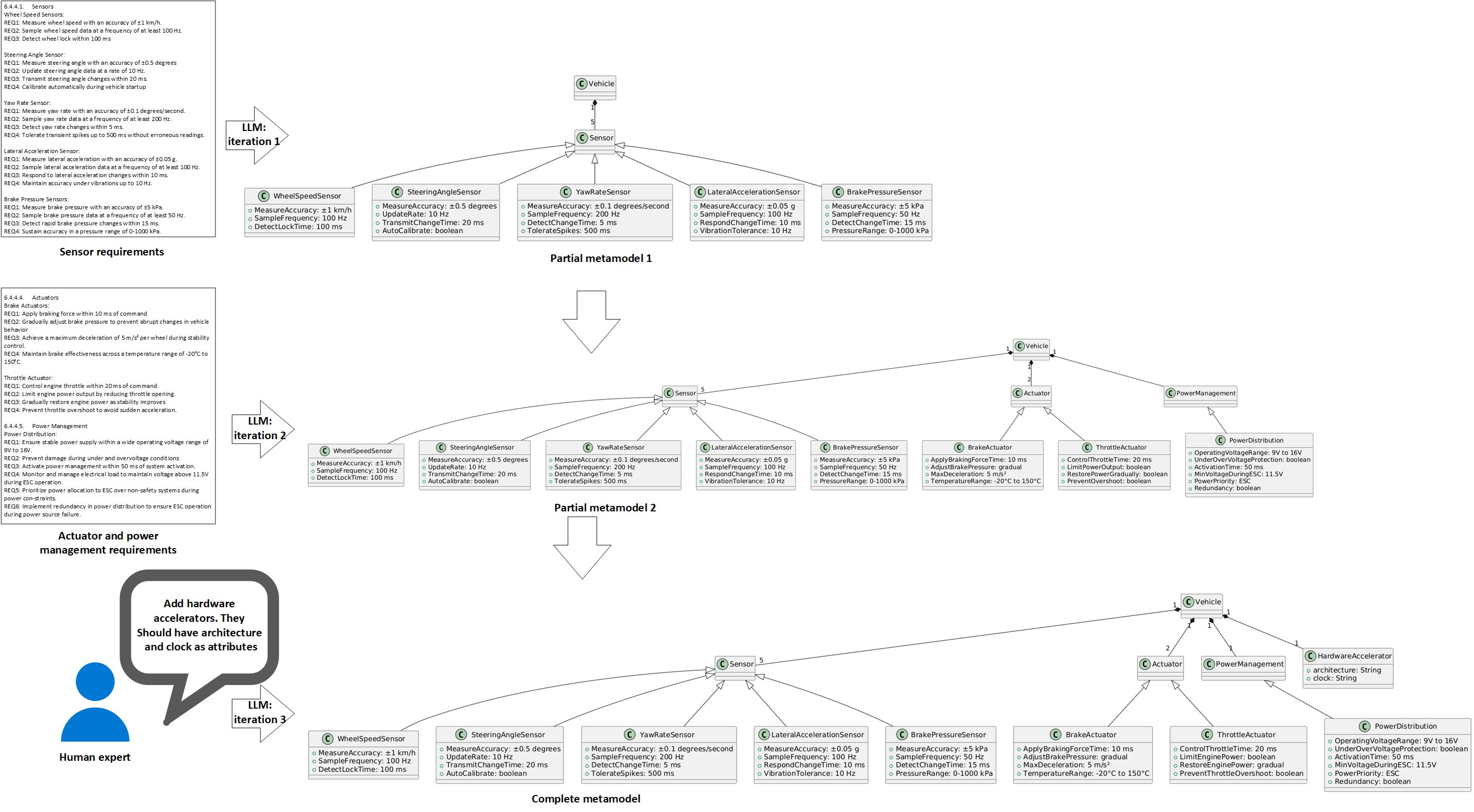}
	\caption{Iterative metamodel construction example: 1) iteration 1-adding sensor requirements 2) iteration 2-adding actuator and power management requirements 3) iteration 3 - human expert feedback}.
	\label{fign2}
\end{figure*}

For purpose of the approach showcase, a scenario of three-iteration automotive metamodel construction using LLM will be depicted (see Fig. \ref{fign2}). 

In the first iteration, we only provide requirements related to sensors. As output, partial metamodel denotes that sensors are part of vehicle, while specific types of sensors with corresponding parameters are derived by inheritance. In the second iteration, we include requirements related to both actuators and power management as well. In the the third iteration, human expert feedback is taken into account, in order to include hardware accelerators as part of the vehicle, considering two relevant properties: processing unit clock and architecture type. The output of the third iteration is complete metamodel containing all the mentioned elements. As it can be noticed, it also includes default attributes values based on the provided OEM requirements.

Additionally, we consider the execution time and token consumption for these iterations, taking into account the number of requirements handled for each of the underlying steps. As it can be seen, both time required and token consumption increase with current metamodel size and number of new requirements introduced. When it comes to the benefits of adopting such approach in automotive toolchains, it is obvious that significant amount of time will be saved, as manual procedure of changing metamodel might be time consuming, especially for huge number of new requirements being abstracted. The estimated effort for such actions done manually would be at least order of magnitude of minute, but automatic approach based on LLMs performs it within several seconds.

\begin{table}[]
\caption{Evaluation for GPT4o solution.}
\begin{tabular}{|l|l|l|l|l|}
\hline
\begin{tabular}[c]{@{}l@{}}Step\end{tabular} & Aspect & Req. [num] & Tokens & Exec. [s] \\ \hline
Creation & Sensors & 19 & 647 & 9.64 \\ \hline
Update & Actuators, power & 14 & 1102 & 11.52 \\ \hline
Feedback & HW accelerators & 3 & 1113 & 9.36 \\ \hline

\end{tabular}
\label{tn1}
\end{table}

\begin{figure*}[]
	\centering
	\includegraphics[width=0.8\textwidth]{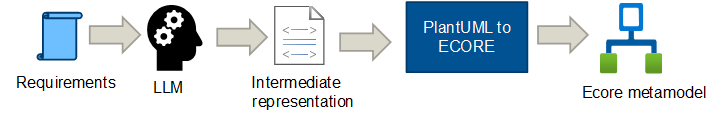}
	\caption{Metamodeling workflow for locally deployable LLM}.
	\label{fign3}
\end{figure*}

On the other side, locally deployable instance of DeepSeek's smaller size model deepseek-ai/deepseek-llm-7b-chat was used for comparison to commercial one. The deployment was done on Google Colab Pro version with A100-based GPU configuration with 40GB of VRAM. Based on the initial results, this model was not able to produce syntactically correct Ecore result. On the other side, it was able to generate correct PlantUML class diagram which can be visualized. Based on that observation, an additional component was added to the pipeline in order to tackle this issue - a model-to-model transformation script which transforms PlantUML class diagrams to Ecore metamodel format, as shown in Fig. \ref{fign3} . The results achieved using locally deployable model were compared to the metamodel constructed using GPT-4o. The summary of the achieved results was given in \ref{tn2}. As it can be noticed, DeepSeek 7B model was able to identify all the classes and composition relations, while its main struggle were generalizations of subclasses. Additionally, it can be also noticed that it exhibits worse performance when it comes to attribute identification, resulting with lower number of attributes in the automatically created product. However, one of possible solutions would be additional steps of human feedback steps to correct the intermediary result.

\begin{table}[]
\caption{Locally deployable DeepSeek 7B solution compared to GPT-4o.}
\begin{tabular}{|l|l|l|l|l|}
\hline
\begin{tabular}[c]{@{}l@{}}Scenario\end{tabular} & Classes & Attributes & Composition & Subclass \\ \hline
Sensors & 6/6 & 15/21 & 5/5 & 6/6 \\ \hline
Actuators & 2/2 & 3/8 & 2/2 & 0/2 \\ \hline
Power & 1/1 & 3/6 & 1/1 & 0/1 \\ \hline

\end{tabular}
\label{tn2}
\end{table}

\section{Conclusion}
In this paper, we have shown a proof-of-concept implementation for iterative metamodeling based on textual input. The presented case study is focused on automotive domain - the inputs are automotive requirements, while the output is metamodel which can be further leveraged for automotive software generation, along with requirements-level completeness and consistency checking. Based on the achieved outcomes, it is shown that such approach exhibits much potential even out of box, without many additional efforts in case of larger commercial models such as OpenAI's GPT-4o, while usage of smaller ones needs adoption of additional strategy in order to ensure the correct and further usable output.

However, several further refinements have to be taken into account, especially to overcome the challenges of handling large metamodels \cite{refn10:Muff2024}, approaches to tackle to LLM hallucinations and cleaning the generated output for usage within automotive tools. Additionally, the potential of locally deployable LLMs will be explored, especially when it comes to smaller models with fine tuning applied, considering that non-disclosure of OEM-specific requirements to external cloud services is highly valuable. Finally, finding the right method and automating the evaluation of LLM-generated metamodels is another challenging aspect that will be addressed in our future works on this topic as well. 

\bibliographystyle{IEEEtran}

\bibliography{biblio}

\begin{thebibliography}{10}
\providecommand{\url}[1]{#1}
\csname url@samestyle\endcsname
\providecommand{\newblock}{\relax}
\providecommand{\bibinfo}[2]{#2}
\providecommand{\BIBentrySTDinterwordspacing}{\spaceskip=0pt\relax}
\providecommand{\BIBentryALTinterwordstretchfactor}{4}
\providecommand{\BIBentryALTinterwordspacing}{\spaceskip=\fontdimen2\font plus
\BIBentryALTinterwordstretchfactor\fontdimen3\font minus
  \fontdimen4\font\relax}
\providecommand{\BIBforeignlanguage}[2]{{%
\expandafter\ifx\csname l@#1\endcsname\relax
\typeout{** WARNING: IEEEtran.bst: No hyphenation pattern has been}%
\typeout{** loaded for the language `#1'. Using the pattern for}%
\typeout{** the default language instead.}%
\else
\language=\csname l@#1\endcsname
\fi
#2}}
\providecommand{\BIBdecl}{\relax}
\BIBdecl

\bibitem{refn1:Petrovic2024}
\BIBentryALTinterwordspacing
N.~Petrovic, F.~Pan, K.~Lebioda, V.~Zolfaghari, S.~Kirchner, N.~Purschke, M.~A.
  Khan, V.~Vorobev, and A.~Knoll, ``{Synergy of Large Language Model and Model
  Driven Engineering for Automated Development of Centralized Vehicular
  Systems},'' Technical University of Munich, Tech. Rep., 2024. [Online].
  Available: \url{https://mediatum.ub.tum.de/doc/1738462/1738462.pdf}
\BIBentrySTDinterwordspacing

\bibitem{refn2:Schulze2012}
M.~Schulze, J.~Weiland, and D.~Beuche, ``Automotive model-driven development
  and the challenge of variability,'' in \emph{Proceedings of the 16th
  International Software Product Line Conference - Volume 1}.\hskip 1em plus
  0.5em minus 0.4em\relax New York, NY, USA: Association for Computing
  Machinery, 2012, p. 207–214.

\bibitem{refn3:Petrovic2023}
N.~Petrovic and I.~Al-Azzoni, ``{Automated Approach to Model-Driven Engineering
  Leveraging ChatGPT and Ecore},'' in \emph{6th International Conference on
  Applied Electromagnetics – PES 2023}, 2023, pp. 166--168.

\bibitem{refn31:Chen2024-1}
K.~Chen, Y.~Yang, B.~Chen, J.~A. Hernández~López, G.~Mussbacher, and
  D.~Varró, ``Automated domain modeling with large language models: A
  comparative study,'' in \emph{2023 ACM/IEEE 26th International Conference on
  Model Driven Engineering Languages and Systems (MODELS)}, 2023, pp. 162--172.

\bibitem{refn32:Chen2024-2}
R.~Chen, J.~Shen, and X.~He, ``{A Model Is Not Built By A Single Prompt:
  LLM-Based Domain Modeling With Question Decomposition},'' 2024.

\bibitem{refn4:Ecore}
\BIBentryALTinterwordspacing
(2025) {Eclipse Ecore Tools}. [Online]. Available:
  \url{https://projects.eclipse.org/projects/modeling.ecoretools}
\BIBentrySTDinterwordspacing

\bibitem{refn5:PlantUML}
\BIBentryALTinterwordspacing
(2025) {PlantUML}. [Online]. Available:
  \url{https://plantweb.readthedocs.io/plantweb/plantweb.render.html}
\BIBentrySTDinterwordspacing

\bibitem{refn6:Petrovic2024-2}
N.~Petrovic, K.~Lebioda, V.~Zolfaghari, A.~Schamschurko, S.~Kirchner,
  N.~Purschke, F.~Pan, and A.~Knoll, ``{LLM-Driven Testing for Autonomous
  Driving Scenarios},'' in \emph{2024 2nd International Conference on
  Foundation and Large Language Models (FLLM)}, 2024, pp. 173--178.

\bibitem{refn7:Zolfaghari2024}
V.~Zolfaghari, N.~Petrovic, F.~Pan, K.~Lebioda, and A.~Knoll, ``{Adopting RAG
  for LLM-Aided Future Vehicle Design},'' in \emph{2024 2nd International
  Conference on Foundation and Large Language Models (FLLM)}, 2024, pp.
  437--442.

\bibitem{refn8:plantweb}
\BIBentryALTinterwordspacing
(2025) {plantweb}. [Online]. Available:
  \url{https://plantweb.readthedocs.io/plantweb/plantweb.html}
\BIBentrySTDinterwordspacing

\bibitem{refn9:Pan2024}
F.~Pan, V.~Zolfaghari, L.~Wen, N.~Petrovic, J.~Lin, and A.~Knoll, ``{Generative
  AI for OCL Constraint Generation: Dataset Collection and LLM Fine-tuning},''
  in \emph{2024 IEEE International Symposium on Systems Engineering (ISSE)},
  2024, pp. 1--8.

\bibitem{refn10:Muff2024}
F.~Muff and H.-G. Fill, ``{Limitations of ChatGPT in Conceptual Modeling:
  Insights from Experiments in Metamodeling},'' {Modellierung 2024 Satellite
  Events}, 2024.

\end{thebibliography}

\balance
\end{document}